\documentclass[fleqn,twoside]{article}
\usepackage[headings]{espcrc2}

\usepackage{axodraw}

\readRCS
$Id: espcrc2.tex,v 1.2 2004/02/24 11:22:11 spepping Exp $
\usepackage{graphicx}
\usepackage[figuresright]{rotating}

\newcommand{\AmS}{{\protect\the\textfont2
  A\kern-.1667em\lower.5ex\hbox{M}\kern-.125emS}}

\title{Unitarity Cuts: NLO Six-Gluon Amplitudes in QCD}

\author{
Pierpaolo Mastrolia\address{
Institut f\"ur Theoretische Physik,Universit\"at Z\"urich\\ 
CH-8057 Z\"urich, Switzerland
}%
\thanks{
based on a joint
project with Ruth Britto and Bo Feng \cite{BFM};
talk given
at Loops \& Legs 2006, 
April 23-28, Eisenach (Germany);
supported
by the European Commission Marie Curie
Fellowship under contract number MEIF-CT-2006-024178.}
}

\runauthor{P. Mastrolia}

\newcommand{\be}{\begin{equation}}
\newcommand{\ee}{\end{equation}}
\newcommand{\nn}{\nonumber}
\newcommand{\bea}{\begin{eqnarray}}
\newcommand{\eea}{\end{eqnarray}}

\def\spa#1.#2{\langle#1\,#2\rangle}
\def\spb#1.#2{[#1\,#2]}

\def\spab#1.#2.#3{\langle\mskip-1mu{#1}
                  | #2 | {#3}]}

\def\spba#1.#2.#3{[\mskip-1mu{#1}
                  | #2 | {#3}\rangle}

\def\spbb#1.#2.#3.#4{[\mskip-1mu{#1}
                     | {#2} \ {#3} | {#4}]}

\def\spaa#1.#2.#3.#4{\langle\mskip-1mu{#1}
                     | {#2} \ {#3} | {#4}\rangle}

\def\dea{\langle \lambda \ d \lambda \rangle}
\def\deb{[\lambda \ d \lambda]}

\def\dedeb{[d \lambda \ \partial_{|\lambda]} ]}
\begin{document}

\begin{abstract}
We report on a technique for evaluating finite unitarity cut
for one-loop amplitudes in gauge theories, and discuss its
application to the cut-constructible part of six-gluon
amplitude in QCD. 
\vspace{1pc}
\end{abstract}

\maketitle

\section{INTRODUCTION}

The availbility of theoretical results required to control 
multi-leg scattering beyond the leading order (LO) 
certainly does not cover the demand for describing 
the landscape 
of multi-particle final-state  processes and backgrounds that
will be delineated at the turn-on of the next generation colliders.
Great efforts over the past 10 years have been made 
to push the status of theoretical results 
beyond processes with five external particles at 
next-to-leading order (NLO) \cite{BernMQ}; 
and we are witnessing nowadays the overpass of the threshold represented by 
six-leg scattering\footnote{see T. Binoth, J. Fujimoto, S. Dittmaier and G. Zanderighi, in this Proceedings}
\cite{Denner:2005es,Ellis:2006ss} 

Evaluation of scattering amplitudes in perturbation theory means computation of 
Feynman diagrams. Currently,
the evaluation of high-tensor-rank one-loop multi-leg 
Feynman integrals represents the bottleneck for NLO calculations. 

One of the most efficient approach to tackle the problem of analytic computation of  
one-loop multileg-amplitudes
is the unitarity cut method proposed by Bern, Dixon, Dunbar and Kosower 
\cite{BernMQ,BernZX,BernCG,BernDB,BernJE}
with the spinor-helicity formalism \cite{berends,xu,gunion}.
The technique hereby described deals with a new way of performing 
the cut-integration, which was 
developed by Britto, Buchbinder, Cachazo and Feng \cite{BrittoHA} in
the context of supersymmetry,
and we have further extended \cite{BFM} 
to deal with the complications due to the tensor structure of propagators
of non-supersymmetric theory, like QCD.
In carring through the integration over the phase-space, we make use of 
'twistor motivated' methods and ideas, 
initiated in the seminal work of Witten \cite{WittenNN} and further developed
in \cite{CachazoKJ,CachazoZB,CachazoBY,CachazoDR}: 
exploiting the properties of analitic continued
amplitudes with complex spinors, to reduce 
the cut-integration to the extraction of residues.

%%%%%%%%%%%%%%%%%%%%%%%%%%%%%%%%%%%%%%%%%%%%%%%%%%%%%%%%%%%%
%%%%%%%%%%%%%%%%%%%%%%%%%%%%%%%%%%%%%%%%%%%%%%%%%%%%%%%%%%%%
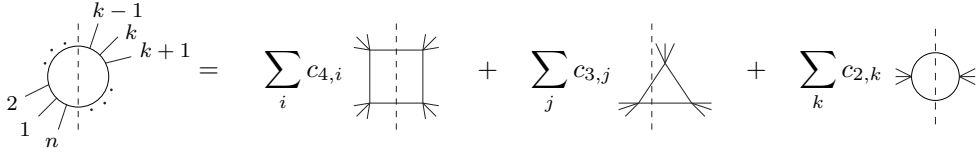
\begin{figure*}[t]
\vspace*{0.5cm}
$$
%%%%%%%%%%%%%%%%%%%%%%%%%%%%%%%%%%%%%%%%%%%%%%%%%%%%%%%
\begin{picture}(0,0)(0,0)
\SetScale{0.5}
\SetWidth{0.5}
\Line(15,40)(5,10)    \Text(15,25)[]{{\footnotesize{$k-1$}}}
\Line(30, 30)(10, 10) \Text(20,17)[]{{\footnotesize{$k$}}}
\Line(40,15)(0,5)     \Text(33,10)[]{{\footnotesize{$k+1$}}}
\Line(-40,-15)(0,5)     \Text(-25,-10)[]{{\footnotesize{$2$}}}
\Line(-30,-30)(-10,-10) \Text(-20,-20)[]{{\footnotesize{$1$}}}
\Line(-15,-40)(-5,-10)  \Text(-10,-25)[]{{\footnotesize{$n$}}}
\GOval(0,0)(23,23)(0){1.0}
\Vertex(-25,+10){1}
\Vertex(-20,+20){1}
\Vertex(-10,+25){1}
\Vertex(+25,-10){1}
\Vertex(+20,-20){1}
\Vertex(+10,-25){1}
\DashLine(0,40)(0,-40){5}
\end{picture}
%%%%%%%%%%%%%%%%%%%%%%%%%%%%%%%%%%%%%%%%%%%%%%%%%%%%%%%
\hspace*{1.5cm} 
=
\hspace*{0.5cm} 
\sum_i c_{4,i} \qquad 
%%%%%%%%%%%%%%%%%%%%%%%%%%%%%%%%%%%%%%%%
\begin{picture}(0,0)(0,0)
\SetScale{0.5}
\SetWidth{0.5}
\Line(-20,-20)(20,-20)
\Line(20,-20)(20,20)
\Line(20,20)(-20,20)
\Line(-20,20)(-20,-20)
\Line(-20,-20)(-32,-22)
\Line(-20,-20)(-30,-30)
\Line(-20,-20)(-22,-32)
\Line(20,-20)(32,-22)
\Line(20,-20)(30,-30)
\Line(20,-20)(22,-32)
\Line(20,20)(32,22)
\Line(20,20)(30,30)
\Line(20,20)(22,32)
\Line(-20,20)(-32,22)
\Line(-20,20)(-30,30)
\Line(-20,20)(-22,32)
\DashLine(0,40)(0,-40){5}
\end{picture}
%%%%%%%%%%%%%%%%%%%%%%%%%%%%%%%%%%%%%%%%
%
\hspace*{1.0cm} 
+
\quad
\sum_j c_{3,j}
\hspace*{0.7cm} 
%
%%%%%%%%%%%%%%%%%%%%%%%%%%%%%%%%%%%%%%%%
\begin{picture}(0,0)(0,0)
\SetScale{0.5}
\SetWidth{0.5}
\Line(-20,-20)(0,10)
\Line(20,-20)(0,10)
\Line(20,-20)(-20,-20)
\Line(0,10)(-7,25)
\Line(0,10)( 0,25)
\Line(0,10)( 7,25)
\Line(-20,-20)(-30,-30)
\Line(-20,-20)(-35,-25)
\Line(-20,-20)(-35,-20)
\Line(20,-20)(30,-30)
\Line(20,-20)(35,-25)
\Line(20,-20)(35,-20)
\DashLine(-10,40)(-10,-40){5}
\end{picture}
%%%%%%%%%%%%%%%%%%%%%%%%%%%%%%%%%%%%%%%%
%
\hspace*{1.0cm} 
+
\quad
\sum_k c_{2,k}
\hspace*{0.7cm} 
%
%%%%%%%%%%%%%%%%%%%%%%%%%%%%%%%%%%%%%%%%
\begin{picture}(0,0)(0,0)
\SetScale{0.6}
\SetWidth{0.6}
\GOval(0,0)(15,15)(0){1}
\Line(-15,0)(-25,+5)
\Line(-15,0)(-25,0)
\Line(-15,0)(-25,-5)
\Line( 15,0)(25,+5)
\Line( 15,0)(25,0)
\Line( 15,0)(25,-5)
\DashLine(0,30)(0,-30){5}
\end{picture}
%%%%%%%%%%%%%%%%%%%%%%%%%%%%%%%%%%%%%%%%
%
$$
\caption{Decomposition of the double-cut of 
a generic one-loop $n$-gluon amplitudes in
  terms of the double-cut of boxes, 
  triangles and bubbles, with rational coefficients $c$'s.}
\label{Deco}
\end{figure*}
%%%%%%%%%%%%%%%%%%%%%%%%%%%%%%%%%%%%%%%%%%%%%%%%%%%%%%%%%%%%
%%%%%%%%%%%%%%%%%%%%%%%%%%%%%%%%%%%%%%%%%%%%%%%%%%%%%%%%%%%%

%%%%%%%%%%%%%%%%%%
\subsection{Setup}
%%%%%%%%%%%%%%%%%%

\subsubsection{Supersymmetry Decomposition}

One-loop amplitudes with all external gluons and a gluon
circulating around the loop, ${\cal A}^{g}$, can be 
rewritten \cite{BernMQ,BernZX,BernCG} by decomposing the inner loop 
as Super-Yang-Mills (SYM) contributions
of a ${\cal N}=4$ and a chiral ${\cal N}=1$ multiplets, 
plus a complex scalar loop, often referred as to ${\cal N}=0$ contribution,
\bea
{\cal A}^{g} = {\cal A}^{{\cal N}=4} 
- 4 {\cal A}^{{\cal N}=1}
+ {\cal A}^{{\cal N}=0} \ .
\label{themiste}
\eea
The main advantage of this decomposition is that supersymmetric
amplitudes, ${\cal A}^{{\cal N}=4}$ and ${\cal A}^{{\cal N}=1}$, are
four-dimensional cut-constructible \cite{BernZX,BernCG}.
Whereas, the term ${\cal A}^{{\cal N}=0}$ contains both a 
polylogarithmic structure
which can be reconstructed from its absorptive part, and a rational remainder 
not detected by 4-dimension cuts.

In \cite{BFM}, we have completed the program of computing the cut-constructible
piece of the NLO six-gluon amplitudes in QCD, by showing a systematic way to evaluate the
cut-constructible piece of ${\cal A}^{{\cal N}=0}$.

\subsubsection{Integral Basis}

By standard reduction techniques, it is known that 
any one-loop gluon amplitude
can be expressed in a basis of scalar integral functions
known as boxes ($I_4$), triangles ($I_3$), and bubbles ($I_2$),
which are analitically available \cite{BernCG,BernKR}.
Indeed, we may exploit the knowledge about its singular behaviour 
\cite{BrittoHA}, 
to express the amplitude in terms of a 
basis of {\it finite} scalar integrals 
(with no one-mass or two-mass triangle functions):
\bea
{\cal A}_{n}^{{\cal N}=0}
&=&
\sum \Big( c_2 I_2 
+ c_3^{3m} I_3^{3m}
+ c_4^{1m} I_{4F}^{1m} 
\nn \\
&& \qquad
+ c_4^{2m~e} I_{4F}^{2m~e} 
+ c_4^{2m~h} I_{4F}^{2m~h} 
\nn \\
&& \qquad
+ c_4^{3m} I_{4F}^{3m} 
+ c_4^{4m} I_{4}^{4m} \Big) \ ,
\label{kole}
\eea
where the subfix $F$ stands for the finite part of the corresponding function.

To compute the amplitude, it is sufficient to compute each of those
coefficients separately and the principle of the unitarity-based
method \cite{BernZX,BernCG,BernDB,BernJE} is to exploit the unitarity
cuts of the scalar integrals to extract their coefficients, see Fig.\ref{Deco}.

\section{THE METHOD (in a nutshell)}

\begin{figure}[hb]
\vspace*{1.0cm}
\hspace*{3cm}
\begin{picture}(0,0)(0,0)
\SetScale{0.8}
\SetWidth{0.8}
\Line(-30,-30)(-10,-10) \Text(-30,-30)[]{{\footnotesize{$i$}}}
\Line(-35,-20)(-10,-5)  \Text(-45,-15)[]{{\footnotesize{$i+1$}}}
\Line(-30, 30)(-10, 10) \Text(-30,30)[]{{\footnotesize{$j$}}}
\Line(-10,-10)(40,-10) \Text(5,-20)[]{{\footnotesize{$l_1$}}}
\Line(-10,10)(40,10)   \Text(5,20)[]{{\footnotesize{$l_2$}}}
\Line(40, 10)(60, 25)   \Text(60,30)[]{{\footnotesize{$j+1$}}}
\Line(40, -10)(60, -25) \Text(60,-30)[]{{\footnotesize{$i-1$}}}
\GOval(-10,0)(20,8)(0){1.0}
\GOval(40,0)(20,8)(0){1.0}
\Vertex(-25,-5){1}
\Vertex(-25,3){1}
\Vertex(-24,10){1}
\Vertex(55,-8){1}
\Vertex(57,0){1}
\Vertex(55,8){1}
\DashLine(15,40)(15,-40){2}
\Text(-8,0)[]{{\footnotesize{$A_L$}}}
\Text(32,0)[]{{\footnotesize{$A_R$}}}
\end{picture}
\vspace*{0.5cm}
\caption{Double-cut in the $P_{ij}$-channel}
\label{gener2cut}
\end{figure}
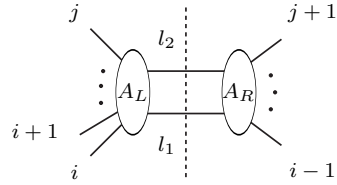

The discontinuity of the amplitude
in the $P=k_i+\ldots+k_j$ momentum channel is computed through the integral
\bea
 C_P = \int d\mu \ 
A^{\rm tree}_L
A^{\rm tree}_R
\label{cutIn}
\eea
with the tree-level amplitudes
$$
\hspace*{-3.0cm}
A^{\rm tree}_L = A^{\rm
  tree}(\ell_1,i,\ldots,j,\ell_2) \ , 
$$
$$
\hspace*{-0.8cm}
A^{\rm tree}_R = A^{\rm
  tree}((-\ell_2),j+1,\ldots ,i-1,(-\ell_1)) 
$$
very efficiently obtained {\it via} BCFW recurrence relation
\cite{BrittoFQ}
and
$$
\hspace*{-0.0cm} d\mu =
d^4\ell_1 d^4\ell_2
\delta^{(+)}(\ell_1^2)\delta^{(+)}(\ell_2^2)\delta^{(4)}
(\ell_1+\ell_2 - P)
$$
being the Lorentz invariant phase-space
measure of two light-like vectors $(\ell_1, \ell_2)$ constrained by
momentum conservation, see Fig.\ref{gener2cut}.

%%%%%%%%%%%%%%%%%%%%%%%%%%%%%%%%%%%%%%%%%%%%%%%%%%%%%%%%%%%%
%%%%%%%%%%%%%%%%%%%%%%%%%%%%%%%%%%%%%%%%%%%%%%%%%%%%%%%%%%%%
%%%%%%%%%%%%%%%%%%%%%%%%%%%%%%%%%%%%%%%%%%%%%%%%%%%%%%%%%%%%
%%%%%%%%%%%%%%%%%%%%%%%%%%%%%%%%%%%%%%%%%%%%%%%%%%%%%%%%%%%%

The four-dimensional delta function allows to integrate one of the propagator
momenta out (say $\ell_2$). 
Then, we rewrite the cut 
measure in terms of spinor variables \cite{CachazoKJ},
\bea
\int d^4 \ell \ \delta^{(+)}(\ell^2) \ 
                \delta^{(+)}((\ell - P)^2) = &&
\nn \\
&&
\hspace*{-5.5cm}
=
\int {\dea \deb \over \spab \lambda.P.\lambda}
\int dt~t~ 
\delta\bigg(t - {P^2 \over \spab \lambda.P.\lambda}\bigg)
\label{meas}
\eea
with $\ell = \ell_1$, and 
where the integration
contour for the spinors is the diagonal 
%$\Bbb{CP}^1$ 
defined by
$\tilde\lambda =\bar\lambda$.

Meanwhile, the integrand, {\it i.e.} the product of the two tree amplitudes 
assumes the generic shape\footnote{
In Eq.(\ref{genform}), consistently with the scaling of the integration measure, 
$\dea \deb$,
the degrees of ${\cal I}$ in both $\lambda$ and $\tilde\lambda$ amount to $-2$.
},
\be
A^{\rm tree}_L A^{\rm tree}_R = \sum_i {\cal I}_i \ ,
\ee
%with,
\be
{\cal I}_i(\ell, \tilde{\ell})
 =
{1 \over \spab \ell.P.\ell^n}
{\spb a_1.\ell \!\ldots\!
 \spa b_1.\ell \!\ldots\!
 \spab \ell.Q_1.\ell \!\ldots
\over
 \spb c_1.\ell \!\ldots\!
 \spa d_1.\ell \!\ldots\!
 \spab \ell.Q_2.\ell \!\ldots
},
\label{genform}
\ee
with $a, b, c, d$'s on-shell spinor, and 
the off-shell vectors $Q$'s $\neq P$.
After rescaling $
|\ell\rangle = \sqrt{t} \ |\lambda\rangle, 
|\ell] = \sqrt{t} \ |\lambda], 
$
one can trivially perform the $t$-integration, 
with the help of the $\delta$ function in Eq.(\ref{meas}).
At this stage, the cut in Eq.(\ref{cutIn}) appears as 
an integral over the spinor variables,
\be
C_P = \sum_i \int \dea \deb \ {\cal I}_i(\lambda, \tilde\lambda)
\label{cut:defIi}
\ee

\subsection{Canonical Decomposition}

In order to perform the spinor integration, it is useful
to write the integrand as a derivative w.r.t. either 
$\lambda$ or $\tilde\lambda$. Without loss of generality, one can decide
for $\tilde\lambda$,
and reduce each term of the integrand to a product of two factors: one
carrying the dependence solely on $\lambda$; 
and a second one, which depends on both, which
will be soon rewritten
as a derivative 
w.r.t. $\tilde\lambda$,
\bea
{\cal I}_i(\lambda, \tilde\lambda) =
{\cal G}_i(\lambda) \ {\cal H}_i(\lambda, \tilde\lambda)
\label{CanDec}
\eea
The decomposition in Eq.(\ref{CanDec}) can be achieved algebraically 
by partial fraction of spinor products, 
using a generalization of the following
Schouten identity
\bea
{ \spb \lambda.{a} \over \spb \lambda.{b} \ \spb \lambda.{c}}
=
  {\spb b.{a} \over \spb b.{c}} {1 \over \spb \lambda.{b}} 
+ {\spb c.{b} \over \spb c.b} {1 \over \spb \lambda.{c}} \ ,
\label{schoutenbra}
\eea
for instance,
\bea
{ \spb \lambda.{a} \over \spba \lambda.{V}.\lambda \spb \lambda.{c}}
=
  {\spab \lambda.V.{a} \over \spab \lambda.V.{c}} {1 \over \spba \lambda.V.{\lambda}} 
+ {\spba c.V.{\lambda} \over \spba c.V.\lambda} {1 \over \spb \lambda.{c}} 
\label{genschoutenbra}
\eea
which holds for any off-shell vector $V$.

The specific shape of ${\cal H}_i(\lambda, \tilde\lambda)$ in
Eq.(\ref{CanDec}) carries 
the {\it signature} of the cut of the polylogarithm which is associated to, and
therefore of its corresponding topology.
In particular we know that:
1) the cut of bubble-functions is rational; 2) the cut of triangle- and
box-functions is logarithmic (with arguments which distinguish unequivocally 
among them).

%%%%%%%%%%%%%%%%%%%%%%%%%%%%%%%%%%%%%%%%%%%%%%%%%%%%%%%%%%%%
%%%%%%%%%%%%%%%%%%%%%%%%%%%%%%%%%%%%%%%%%%%%%%%%%%%%%%%%%%%%
\begin{figure*}[t]
\begin{eqnarray}
\hspace*{1.0cm}
%%%%%%%%%%%%%%%%%%%%%%%%%%%%%%%%%%%%%%%%
%%%%%%%%%%%%%%%%%%%%%%%%%%%%%%%%%%%%%%%%
%%%%%%%%%%%%%%    a   %%%%%%%%%%%%%%%%%%
%%%%%%%%%%%%%%%%%%%%%%%%%%%%%%%%%%%%%%%%
\begin{picture}(0,0)(0,0)
\SetScale{0.35}
\SetWidth{0.35}
\Line(35,20)(0,35)
\Line(0,35)(-35,20)
\Line(-35,20)(-35,-20)
\Line(-35,-20)(0,-35)
\Line(0,-35)(35,-20)
\Line(35,-20)(35,20)
\Line(35,20)(55,30)
\Line(-35,20)(-55,30)
\Line(-35,-20)(-55,-30)
\Line(35,-20)(55,-30)
\Line(0,35)(0,50)
\Line(0,-35)(0,-50)
\DashLine(30,45)(0,0){5}
\DashLine(-50,0)(0,0){5}
\DashLine(-30,-45)(0,0){5}
\DashLine(30,-45)(0,0){5}
\end{picture}
%%%%%%%%%%%%%%%%%%%%%%%%%%%%%%%%%%%%%%%%
\hspace*{0.8cm}
\to
\hspace*{0.8cm}
%%%%%%%%%%%%%%%%%%%%%%%%%%%%%%%%%%%%%%%%
\begin{picture}(0,0)(0,0)
\SetScale{0.5}
\SetWidth{0.5}
\Line(-20,-20)(20,-20)
\Line(20,-20)(20,20)
\Line(20,20)(-20,20)
\Line(-20,20)(-20,-20)
\Line(-20,-20)(-30,-30)
\Line(20,-20)(30,-30)
\Line(20,20)(32,22)
\Line(20,20)(22,32)
\Line(-20,20)(-32,22)
\Line(-20,20)(-22,32)
\DashLine(-30,0)(-10,0){5}
\DashLine( 10,0)( 30,0){5}
\DashLine(0,30)(0,10){5}
\DashLine(0,-10)(0,-30){5}
\end{picture}
%%%%%%%%%%%%%%%%%%%%%%%%%%%%%%%%%%%%%%%%
%%%%%%%%%%%%%%%%%%%%%%%%%%%%%%%%%%%%%%%%
%%%%%%%%%%%%%%    b   %%%%%%%%%%%%%%%%%%
%%%%%%%%%%%%%%%%%%%%%%%%%%%%%%%%%%%%%%%%
\hspace*{3.0cm}
%%%%%%%%%%%%%%%%%%%%%%%%%%%%%%%%%%%%%%%%
\begin{picture}(0,0)(0,0)
\SetScale{0.35}
\SetWidth{0.35}
\Line(35,20)(0,35)
\Line(0,35)(-35,20)
\Line(-35,20)(-35,-20)
\Line(-35,-20)(0,-35)
\Line(0,-35)(35,-20)
\Line(35,-20)(35,20)
\Line(35,20)(55,30)
\Line(-35,20)(-55,30)
\Line(-35,-20)(-55,-30)
\Line(35,-20)(55,-30)
\Line(0,35)(0,50)
\Line(0,-35)(0,-50)
\DashLine(30,45)(0,0){5}
\DashLine(-50,0)(0,0){5}
\DashLine(-30,-45)(0,0){5}
\DashLine(50,0)(0,0){5}
\end{picture}
%%%%%%%%%%%%%%%%%%%%%%%%%%%%%%%%%%%%%%%%
\hspace*{0.8cm}
\to
\hspace*{0.8cm}
%%%%%%%%%%%%%%%%%%%%%%%%%%%%%%%%%%%%%%%%
\begin{picture}(0,0)(0,0)
\SetScale{0.5}
\SetWidth{0.5}
\Line(-20,-20)(20,-20)
\Line(20,-20)(20,20)
\Line(20,20)(-20,20)
\Line(-20,20)(-20,-20)
%
%% \Line(-20,-20)(-32,-22)
\Line(-20,-20)(-30,-30)
%% \Line(-20,-20)(-22,-32)
%
\Line(20,-20)(32,-22)
%% \Line(20,-20)(30,-30)
\Line(20,-20)(22,-32)
%
%% \Line(20,20)(32,22)
\Line(20,20)(30,30)
%% \Line(20,20)(22,32)
%
\Line(-20,20)(-32,22)
%% \Line(-20,20)(-30,30)
\Line(-20,20)(-22,32)
\DashLine(-30,0)(-10,0){5}
\DashLine( 10,0)( 30,0){5}
\DashLine(0,30)(0,10){5}
\DashLine(0,-10)(0,-30){5}
\end{picture}
%%%%%%%%%%%%%%%%%%%%%%%%%%%%%%%%%%%%%%%%
%%%%%%%%%%%%%%%%%%%%%%%%%%%%%%%%%%%%%%%%
%%%%%%%%%%%%%%%%%%%%%%%%%%%%%%%%%%%%%%%%
%%%%%%%%%%%%%%    c   %%%%%%%%%%%%%%%%%%
%%%%%%%%%%%%%%%%%%%%%%%%%%%%%%%%%%%%%%%%
\hspace*{3.0cm}
%%%%%%%%%%%%%%%%%%%%%%%%%%%%%%%%%%%%%%%%
\begin{picture}(0,0)(0,0)
\SetScale{0.35}
\SetWidth{0.35}
\Line(35,20)(0,35)
\Line(0,35)(-35,20)
\Line(-35,20)(-35,-20)
\Line(-35,-20)(0,-35)
\Line(0,-35)(35,-20)
\Line(35,-20)(35,20)
\Line(35,20)(55,30)
\Line(-35,20)(-55,30)
\Line(-35,-20)(-55,-30)
\Line(35,-20)(55,-30)
\Line(0,35)(0,50)
\Line(0,-35)(0,-50)
\DashLine(30,45)(0,0){5}
\DashLine(-30,-45)(0,0){5}
\DashLine(30,-45)(0,0){5}
\DashLine(50,0)(0,0){5}
\end{picture}
%%%%%%%%%%%%%%%%%%%%%%%%%%%%%%%%%%%%%%%%
\hspace*{0.8cm}
\to
\hspace*{0.8cm}
%%%%%%%%%%%%%%%%%%%%%%%%%%%%%%%%%%%%%%%%
%%%%%%%%%%%%%%%%%%%%%%%%%%%%%%%%%%%%%%%%
\begin{picture}(0,0)(0,0)
\SetScale{0.5}
\SetWidth{0.5}
\Line(-20,-20)(20,-20)
\Line(20,-20)(20,20)
\Line(20,20)(-20,20)
\Line(-20,20)(-20,-20)
%
%% \Line(-20,-20)(-32,-22)
\Line(-20,-20)(-30,-30)
%% \Line(-20,-20)(-22,-32)
%
%% \Line(20,-20)(32,-22)
\Line(20,-20)(30,-30)
%% \Line(20,-20)(22,-32)
%
%% \Line(20,20)(32,22)
\Line(20,20)(30,30)
%% \Line(20,20)(22,32)
%
\Line(-20,20)(-32,22)
\Line(-20,20)(-30,30)
\Line(-20,20)(-22,32)
\DashLine(-30,0)(-10,0){5}
\DashLine( 10,0)( 30,0){5}
\DashLine(0,30)(0,10){5}
\DashLine(0,-10)(0,-30){5}
\end{picture}
%%%%%%%%%%%%%%%%%%%%%%%%%%%%%%%%%%%%%%%%
%
\nn
\end{eqnarray}
\caption{Quadruple-cuts of a one-loop six-point amplitude are associated to
  box-function coefficients.}
\label{Q4cuts}
\end{figure*}
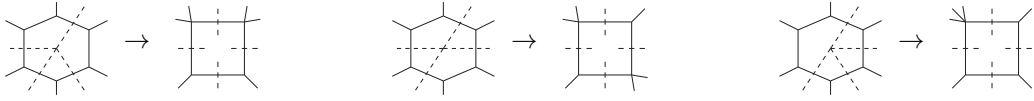
%%%%%%%%%%%%%%%%%%%%%%%%%%%%%%%%%%%%%%%%%%%%%%%%%%%%%%%%%%%%
%%%%%%%%%%%%%%%%%%%%%%%%%%%%%%%%%%%%%%%%%%%%%%%%%%%%%%%%%%%%

%%%%%%%%%%%%%%%%%%%%%%%%%%%%%%%%%%%%%%
\subsection{Rational terms of the cut}
%%%%%%%%%%%%%%%%%%%%%%%%%%%%%%%%%%%%%%
One possibility is: 
\bea
{\cal H}_i(\lambda, \tilde\lambda) = 
{ \spb \eta.\lambda^n \over \spab \lambda.P.\lambda^{n+2}} \ .
\eea 
In this case one can perform the integration over the $\tilde\lambda$-variable
by parts,
\bea
{\deb \ \spb \eta.\lambda^n \over \spab \lambda.P.\lambda^{n+2}} 
=
{\dedeb \over (n+1)}
{
\spb \eta.\lambda^{n+1} 
\over \spab \lambda.P.\lambda^{n+1} \spab \lambda.P.\eta} 
\label{ibp}
\eea

\subsubsection{Cauchy residue theorem}

Finally, one performs the last integration over the $\lambda$-variable 
using Cauchy residue theorem, in the fashion of the holomorphic anomaly
\cite{CachazoBY,CachazoDR}, according to which the following relation does
hold,
\bea
\dedeb {1 \over \spa\lambda.a} 
= d\tilde{\lambda}^{\dot{a}}
         {\partial \over \partial {\lambda}^{\dot{a}}}
         {1 \over \spa\lambda.a} 
= 2 \pi \delta(\spa\lambda.a)
\eea
Therefore the contribution to Eq.(\ref{cut:defIi}) 
reads,
\bea
\int \dea \deb \ {\cal I}_i(\lambda, \tilde\lambda) = && 
\nn \\
= \int {\dea \dedeb  \over (n+1)}
{
{\cal G}_i(\lambda) \ 
\spb \eta.\lambda^{n+1} 
\over 
\spab \lambda.P.\lambda^{n+1} 
\spab \lambda.P.\eta
} 
= && \nn \\
&&
\hspace*{-7.0cm} 
= {1 \over (n+1)} \bigg\{
{
{\cal G}_i(\lambda)
\over 
(P^2)^{n+1}
}\bigg|_{|\lambda\rangle = P|\eta]}
\nn \\
&&
\hspace*{-6.0cm}
+
\sum_j \lim_{\lambda \to \lambda_{ij}} 
\spa \lambda_{ij}.\lambda
{
{\cal G}_i(\lambda) \ 
\spb \eta.\lambda^{n+1} 
\over 
\spab \lambda.P.\lambda^{n+1} 
\spab \lambda.P.\eta
}
\bigg\} 
\label{Cauchy}
\eea
where $|\lambda_{ij} \rangle$ are the simple poles of ${\cal G}_i(\lambda)$.
Since the final result is a rational number, it can be taken as 
the coefficient of the finite cut of a bubble-function with external momentum
$P$. 

%%%%%%%%%%%%%%%%%%%%%%%%%%%%%%%%%%%%%%%%
\subsection{Logarithmic terms of the cut}
%%%%%%%%%%%%%%%%%%%%%%%%%%%%%%%%%%%%%%%%
The other possibility for the expression of ${\cal H}_i$ can be represented by,
\bea
{\cal H}_i(\lambda, \tilde\lambda) = 
{ 
1
\over 
\spab \lambda.Q_1.\lambda
\spab \lambda.Q_2.\lambda
} \ .
\eea 

\subsubsection{Feynman parameterization}
In this case, to walk along the same path as in the previous section,
we have to introduce a Feynman parameter and write, 
\bea
{\cal H}_i(\lambda, \tilde\lambda) &=& 
\int_0^1 d x \  
{ 
1
\over 
\spab \lambda.R.\lambda^2
} \ ,
\\
R = R(x) &=& x Q_1 + (1-x) Q_2 \ ,
\eea 
so that the whole integral becomes
\bea
\int_0^1 d x 
\int \dea \deb \ {{\cal G}_i(\lambda) \over \spab \lambda.R.\lambda^2} \ .
\label{I2}
\eea

\subsubsection{Cauchy residue theorem}

Having the integrand in this form,
one can proceed integrating-by-parts over the $\tilde\lambda$-variable
with the help of Eq.(\ref{ibp}), and over the $\lambda$-variable
using Cauchy residue theorem as in Eq.(\ref{Cauchy}) (setting $n=0$), 
obtaining
\bea
&&
%\hspace*{-5.7cm}
\int_0^1 dx 
\bigg\{
{
{\cal G}_i(\lambda)
\over 
(R^2)
}\bigg|_{|\lambda\rangle = R|\eta]}
\nn \\
&&
%\hspace*{-5.5cm}
+
\sum_j \lim_{\lambda \to \lambda_{ij}} 
\spa \lambda_{ij}.\lambda
{
{\cal G}_i(\lambda) \ 
\spb \eta.\lambda 
\over 
\spab \lambda.R.\lambda
\spab \lambda.R.\eta
} 
\bigg\}
\eea

\subsubsection{Feynman integration}
The left over integration over the Feynman parameter is the source of the
logarithmic part.
Since $R^2$ is quadratic in $x$, it can be written as,
\bea
R^2 = (x - x_1) (x - x_2) \ ,
\eea
with $x_{1,2}$ being the two real roots of the quadratic equation $R^2 = 0$.
The term ${\cal G}_i(\lambda)$ at $|\lambda\rangle = R|\eta]$ is instead a
 sum of ratios of polynomial\footnote{
Notice that in Eq.(\ref{I2}),
${\cal G}_i(\lambda)$ is of degree zero in $\lambda$, therefore it
can be written as a combination of terms like
${\spa \eta_1.\lambda^\alpha \over \spa \eta_2.\lambda^\alpha}$, 
with $\alpha \ge 0$, and generic spinors $\eta$'s. 
}  in $x$, therefore the first term in the above integrand will behave like,
\bea
{
{\cal G}_i(\lambda)
\over 
(R^2)
}\bigg|_{|\lambda\rangle = R|\eta]}
\sim
{(1 + \xi_1 x)^\alpha \over (1 + \xi_2 x)^\alpha} {1 \over (x - x_1) (x - x_2)}
\eea
with $\alpha \ge 0$, and $\xi_1, \xi_2$ being ratios of spinor products.
Carrying out the integration over the Feynman parameter does generate indeed
logarithms of the form $\ln(f(x_1,x_2))$. The explicit expression of the
argument is the {\it character} of the corresponding cut-function: 
if $f$ is rational (in the Mandelstam invariants), 
then it is a signature of 1m-, 2m- and 3m-box function;
else if $f$ is irrational, then it is a signature of either 3m-triangle or 
4m-box function.
As a matter of fact one encounters triangle-functions when $Q_1 = P$ and $Q_2
\ne P$; while box-functions arise when $Q_1 \ne Q_2 \ne P$.

%%%%%%%%%%%%%%%%%%%%%%%%%%%%%%%%%%%%%%
\subsection{Treatment of higher poles}
%%%%%%%%%%%%%%%%%%%%%%%%%%%%%%%%%%%%%%
In applying Cauchy residue theorem in Eq.(\ref{Cauchy}), we have implicitly
assumed the integrand as having only {\it simple} poles.
That is indeed the case for scattering amplitudes in ${\cal N}=4,1$ SYM. But certainly
not, within the framework of less supersymmetry, like in QCD.
The problem of lifting up the simple poles hidden beneath the higher poles can
be tackled algebraically, as well. 
%- without taking derivative w.r.t. spinor
%variables, as theory of complex function would suggest at first sight.
In fact, one can iteratively use the parity-conjugated 
($\spb{}.{} \leftrightarrow \spa{}.{}$) of Schouten identities
Eqs.(\ref{schoutenbra},\ref{genschoutenbra}), 
for partial fractioning the spinor products in the denominator.
The effect is an algebraic 
disentangling of the poles, and allows to
give the integrand, which carries a degree -2 in $\lambda$, the following shape
$$
{g_{i,-1}(\lambda, \tilde{\lambda}) 
\over \spa \lambda.\omega}
+ \sum_{j=2}^\beta 
{ 
\spa \lambda.{\phi}^{j-2} 
\over 
\spa \lambda.{\omega}^j 
}
g_{i,j}(\tilde{\lambda})
$$
with $g_{i,-1}(\lambda, \tilde{\lambda})$ having degree -1 in $\lambda$, and at most
{\it simple} poles in $|\lambda\rangle \ne \omega$; and 
$g_{i,j}(\tilde{\lambda})$ independent of $\lambda$.
Therefore, when the index of the sum over $j$ in Eq.(\ref{Cauchy}) hits
the higher pole $\lambda_{ij} = \omega$, only the 
first term in the above expression will have the non-vanishing
residue,
$
g_{i,-1}(\lambda=\omega, \tilde{\lambda}=\bar{\omega}).
$

%%%%%%%%%%%%%%%%%%%%%%%%%%%%%%%%%%%%%%%%%%%%%%%%%%%%%%%%%%%%
%%%%%%%%%%%%%%%%%%%%%%%%%%%%%%%%%%%%%%%%%%%%%%%%%%%%%%%%%%%%
\begin{figure}
\begin{eqnarray}
\hspace*{1.0cm}
%%%%%%%%%%%%%%%%%%%%%%%%%%%%%%%%%%%%%%%%
\begin{picture}(0,0)(0,0)
\SetScale{0.35}
\SetWidth{0.35}
\Line(35,20)(0,35)
\Line(0,35)(-35,20)
\Line(-35,20)(-35,-20)
\Line(-35,-20)(0,-35)
\Line(0,-35)(35,-20)
\Line(35,-20)(35,20)
\Line(35,20)(55,30)
\Line(-35,20)(-55,30)
\Line(-35,-20)(-55,-30)
\Line(35,-20)(55,-30)
\Line(0,35)(0,50)
\Line(0,-35)(0,-50)
\DashLine(30,45)(-30,-45){5}
\end{picture}
%%%%%%%%%%%%%%%%%%%%%%%%%%%%%%%%%%%%%%%%
\hspace*{0.8cm}
&\to& 
\hspace*{0.8cm}
%
%%%%%%%%%%%%%%%%%%%%%%%%%%%%%%%%%%%%%%%%
\begin{picture}(0,0)(0,0)
\SetScale{0.6}
\SetWidth{0.6}
\GOval(0,0)(15,15)(0){1}
\Line(-15,0)(-25,+5)
\Line(-15,0)(-25,0)
\Line(-15,0)(-25,-5)
\Line( 15,0)(25,+5)
\Line( 15,0)(25,0)
\Line( 15,0)(25,-5)
\DashLine(0,30)(0,-30){5}
\end{picture}
%%%%%%%%%%%%%%%%%%%%%%%%%%%%%%%%%%%%%%%%
\nn \\
&& \nn \\
&& \nn \\
%%%%%%%%%%%%%%%%%%%%%%%%%%%%%%%%%%%%%%%%
%%%%%%%%%%%%%%%%%%%%%%%%%%%%%%%%%%%%%%%%
%%%%%%%%%%%%%%%%%%%%%%%%%%%%%%%%%%%%%%%%
\begin{picture}(0,0)(0,0)
\SetScale{0.35}
\SetWidth{0.35}
\Line(35,20)(0,35)
\Line(0,35)(-35,20)
\Line(-35,20)(-35,-20)
\Line(-35,-20)(0,-35)
\Line(0,-35)(35,-20)
\Line(35,-20)(35,20)
\Line(35,20)(55,30)
\Line(-35,20)(-55,30)
\Line(-35,-20)(-55,-30)
\Line(35,-20)(55,-30)
\Line(0,35)(0,50)
\Line(0,-35)(0,-50)
\DashLine(-20,45)(-20,-45){5}
\end{picture}
%%%%%%%%%%%%%%%%%%%%%%%%%%%%%%%%%%%%%%%%
\hspace*{0.8cm}
&\to& 
\hspace*{0.82cm}
%%%%%%%%%%%%%%%%%%%%%%%%%%%%%%%%%%%%%%%%
\begin{picture}(0,0)(0,0)
\SetScale{0.6}
\SetWidth{0.6}
\GOval(0,0)(15,15)(0){1}
\Line(-15,0)(-30,+7)
%\Line(-15,0)(-25,0)
\Line(-15,0)(-30,-7)
\Line( 15,0)(27,+15)
\Line( 15,0)(30,+5)
\Line( 15,0)(30,-5)
\Line( 15,0)(27,-15)
\DashLine(0,30)(0,-30){5}
\end{picture}
%%%%%%%%%%%%%%%%%%%%%%%%%%%%%%%%%%%%%%%%
%
\hspace*{1.0cm} 
\&
\hspace*{0.8cm} 
%
%%%%%%%%%%%%%%%%%%%%%%%%%%%%%%%%%%%%%%%%
\begin{picture}(0,0)(0,0)
\SetScale{0.5}
\SetWidth{0.5}
\Line(-20,-20)(0,10)
\Line(20,-20)(0,10)
\Line(20,-20)(-20,-20)
\Line(0,10)(-7,25)
%% \Line(0,10)( 0,25)
\Line(0,10)( 7,25)
\Line(-20,-20)(-30,-30)
%% \Line(-20,-20)(-35,-25)
\Line(-20,-20)(-35,-20)
\Line(20,-20)(30,-30)
%% \Line(20,-20)(35,-25)
\Line(20,-20)(35,-20)
\DashLine(-10,30)(-10,-40){5}
\end{picture}
%%%%%%%%%%%%%%%%%%%%%%%%%%%%%%%%%%%%%%%%
%
\nn
\end{eqnarray}
\caption{Double-cuts of a one-loop six-point amplitude contain bubble- and
  3m-triangle coefficients (and boxes' which are not shown).}
\label{D2cuts}
\end{figure}
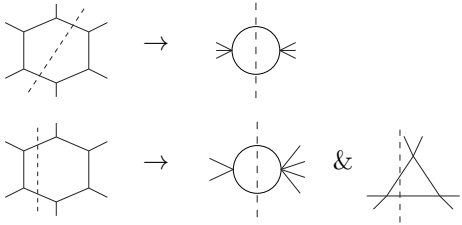
%%%%%%%%%%%%%%%%%%%%%%%%%%%%%%%%%%%%%%%%%%%%%%%%%%%%%%%%%%%%
%%%%%%%%%%%%%%%%%%%%%%%%%%%%%%%%%%%%%%%%%%%%%%%%%%%%%%%%%%%%

\subsection{Out of the Cut }

To summarize, every double-cut of a given amplitude does contain box-,
triangle- and bubble-cut, each multiplied by its own coefficient,
as depicted in Fig.\ref{Deco}.
The {\it canonical decomposition} disentangles in a purely algebraic way the
rational terms from the logarithmic terms: the former are associated to the
cut of two-point functions; the latter, to the cut of three- and four-point
functions. The argument of each logarithm specifies unequivocally the topology
of the functions it is associated to. Alternatively, the coefficient of the
box-function can be algebraically determined by freezing the loop momenta with the
four conditions imposed by quadruple-cutting the amplitude, as extensively
discussed in \cite{BrittoNC}.

%%%%%%%%%%%%%%%%%%%%%%%%%%%%%%%%%
\section{NLO Six-Gluon Amplitude}
%%%%%%%%%%%%%%%%%%%%%%%%%%%%%%%%%

The helicity amplitudes for the six-gluon scattering have been recently
computed {\it via} semi-numerical technique\footnote{see G. Zanderighi, in
  this Proceedings} 
\cite{Ellis:2006ss}. 
The status of the efforts in computing them analytically is reported in
Tab. \ref{table:status6}. 
The completion of the cut-constructible piece required the computation of
the three-minus ${\cal N}=0$ contributions:
the coefficients of box-functions had already been computed {\it via} quadruple-cuts
\cite{BidderRI}, see in Fig.\ref{Q4cuts}; 
the answer for the cut-constructible $A(---+++)$
has been obtained by recursive technique\footnote{see D. Dunbar, in
  this Proceedings}  
\cite{BernHH}; 
while we have computed \cite{BFM} 
{\it via} double cuts, see Fig.\ref{D2cuts},
the coefficients of bubbles- and 3m-triangles of $A(--+-++)$ and $A(-+-+-+)$ 
according to the method discussed in the previous section.

The completion of the analytical result demands the computation of the
rational term, not detected by the cuts in 4-dimension.
On one side, the exploitation of properties like 
soft-collinear singularities and 
factorization, and the knowledge of the polylogarithmic terms
(which are cut constructible), on the other side,
have lead to uncover the recursive behavior of the rational term
of one-loop gluon amplitude \cite{BernJI,FordeHH,BernCQ,Berger:2006ci}. 
Thanks to this progress, the rational terms of  
$A(--++++)$ and $A(---+++)$ 
have been recently obtained by the bootstrap method
\cite{BernCQ,Berger:2006ci}\footnote{see Z. Bern and D. Kosower, in
  this Proceedings}, 
and the rational contribution coming from the other helicity configurations is
at the horizon.

\begin{table}[tb]
\caption{Status of the analytic computation of the one-loop correction to the
  six-gluon amplitude}
\label{table:status6}
\begin{tabular}{|c||c|c|c|c|}
\hline
Amplitude  &$\!\!{\cal N}\!=\!4\!\!$& 
            $\!\!{\cal N}\!=\!1\!\!$ & 
            $\!\!{\cal N}\!=\!0\!\!$ &
            $\!\!{\cal N}\!=\!0\!\!$ \\
           &  & 
              & cut
              & rat \\
\hline
\hline
$- - + + + +$ 
           & \cite{BernZX}
           & \cite{BernCG}
           & \cite{BernCG}
           & \cite{BernCQ}
\\
\hline
$- + - + + +$ 
           & \cite{BernZX}
           & \cite{BernCG}
           & \cite{BedfordNH}
           & 
\\
\hline
$- + + - + +$ 
           & \cite{BernZX}
           & \cite{BernCG}
           & \cite{BedfordNH}
           & 
\\
\hline
$- - - + + +$
           & \cite{BernCG}
           & \cite{BidderTX}
           & \cite{BernHH,BFM}
           & \cite{Berger:2006ci}
\\
\hline
$- - + - + +$ 
           & \cite{BernCG}
           & $\!\!$\cite{BrittoHA,BidderVX,BidderRI}$\!\!$
           & \cite{BFM}
           & 
\\
\hline
$- + - + - +$ 
           & \cite{BernCG}
           & $\!\!$\cite{BrittoHA,BidderVX,BidderRI}$\!\!$
           & \cite{BFM}
           & 
\\
\hline
\end{tabular}
\end{table}

\section{Squeezing Out of the Bottleneck}

The importance of the method here outlined \cite{BrittoHA,BFM} 
is that it is a general method for
computing finite cuts of one-loop multi-leg amplitudes. 
It has the non-trivial advantage of not encountering at all
the main difficulties which arise from the
standard tensor reduction;  
and the computational problem is algebraically reduced by trivial spinor algebra to the 
the extraction of residues. 
Such a method is therefore suitable for cut-constructible amplitudes, 
for instance in Super-Yang-Mills and Gravity\footnote{see N.E.J. Bjerrum-Bohr, in
  this Proceedings}\cite{Bern:2005bb},   
and to be used in ping-pong with techniques 
like the bootstrap method \cite{BernJI,FordeHH,BernCQ,Berger:2006ci}, or any
other one which could provide the reconstruction of the rational remainder. 

The six-point amplitudes obtained in \cite{BFM} contain the complete
polylogarithm structure of the 
all-$n$ gluon amplitude at NLO 
(except for $I_4^{4m}$ which can be anyhow computed {\it via}
4ple-cut),
therefore it could be used as bootstrap point to tackle problems with the
higher number of legs - once the recursive behavior of loop amplitudes will
be completely sorted out.

\vspace*{0.5cm}
\noindent
{\bf Acknowledgments}

I whish to thank the invaluable collaboration of Ruth Britto and Bo Feng, and
the very stimulating discussions with Babis Anastasiou and Zoltan Kunszt.

\end{document}